\documentclass{article}

\usepackage{amsmath}

\usepackage{oljour}
\usepackage{amssymb}
\usepackage{graphicx}
\usepackage{dcolumn}
\usepackage{bm}

\newcommand{\cm}{cm$^{-1}$} \newcommand{\A}{\AA$^{-1}$} \newcommand{\Q}{\mathbf{Q}}

\begin{document}


\journalname{zp}  


\title[en]{Where are protons and deuterons in KH$_{p}$D$_{1-p}$CO$_3$? A neutron diffraction study }

\begin{author}
  \anumber{1}   
  \atitle{Dr}    
  \firstname{Fran\c{c}ois}  
  \surname{Fillaux}    
  \vita{}      
  \institute{CNRS-LADIR, UPMC, Univ Paris 06, UMR7075, F-94320, Thiais, France }   
  \street{rue H. Dunant}    
  \number{2}    
  \zip{94320}      
  \country{France}   
  \tel{+33149781283}      
  \fax{+33149781118}      
  \email{fillaux@glvt-cnrs.fr}     
\end{author}

\begin{author}
  \anumber{2}   
  \atitle{Dr}    
  \firstname{Alain}  
  \surname{Cousson}    
  \vita{}      
  \institute{Laboratoire L\'{e}on Brillouin (CEA-CNRS) C.E. Saclay}   
  \street{}    
  \number{}    
  \zip{91191}      
  \country{France}   
  \tel{}      
  \fax{}      
  \email{Cousson@llb.saclay.cea.fr}     
\end{author}

\corresponding{LADIR-CNRS, 2 rue H. Dunant, F-94320 Thiais, France}

\abstract{The crystals of potassium hydrogen carbonate (KHCO$_3$) and the KDCO$_3$ analogue are isomorphous. They are composed of hydrogen or deuterium bonded centrosymmetric dimers (HCO$_3^-)_2$ or (DCO$_3^-)_2$. The space group symmetry of KH$_{p}$D$_{1-p}$CO$_3$ ($p \approx 0.75)$ determined with neutron diffraction is identical to those of KHCO$_3$ and KDCO$_3$. This is at variance with a random distribution of H and D nuclei. These crystals are macroscopic quantum systems in which protons or/and deuterons merge into macroscopic states. }

\zusammenfassung{}


\schlagwort{}

\dedication{It is a great pleasure to dedicate this paper to Professor Hans-Heinrich Limbach on the occasion of his sixty fifth birthday, and to wish him many more happy and productive years of activity in chemistry. }

\received{}
\accepted{}
\volume{}
\issue{}
\class{}
\Year{}

\maketitle

\section*{Introduction}

The formalism of quantum mechanics extrapolated from the level of electrons and atoms to that of everyday life leads to conclusions totally alien to commonsense, such as Schr\"{o}dinger's Cat in a superposition of ``alive-dead'' states and nonlocal observables \cite{Laloe,Leggett}. Such conflicts lead to a dichotomy of interpretation consisting in that, while at the microscopic level a quantum superposition indicates a lack of definiteness of outcome, at the macroscopic level a similar superposition can be interpreted as simply a measure of the probability of one outcome or the other, one of which is definitely realized for each measurement of the ensemble \cite{Leggett1,LG,Leggett5}. For open systems, this dichotomy can be legitimated by decoherence \cite{Zurek} stipulating that an initial superposition state should lose its ability to exhibit quantum interferences via interaction with the environment. However, since the quantum theory does not predict any definite dividing line between quantal and classical regimes, macroscopic quantum behaviour is possible for systems decoupled from, or very weakly coupled to, the surroundings \cite{ACAL}. In principle, there is no upper limit in size, complexity, and temperature, beyond which such systems should be doomed to classicality. 

As a matter of fact, defect-free crystals are macroscopic quantum systems with discrete phonon states at any temperature below melting or decomposition. This is an unavoidable consequence of the translational invariance of the lattice. Yet, the dichotomy of interpretation arises for O--H$\cdots$O hydrogen bonds, when the coexistence of two configurations, say $\mathrm{O1-H}\cdots \mathrm{O2}$ and $\mathrm{O1}\cdots \mathrm{H-O2}$, is conceived of as ``disorder'' \cite{SZS}. Protons are thought of as dimensionless particles, with definite positions and momenta, moving in a double-well coupled to an incoherent thermal bath. By contrast, vibrational spectra provide unquestionable evidences of phonon states showing that the translational invariance and the quantum nature of lattice dynamics are not destroyed by proton transfer. It is therefore compulsory to elaborate a purely quantum rationale avoiding any classical ingredient \cite{Fil7}.

Such a theoretical framework has been elaborated for the crystal of potassium hydrogen carbonate (KHCO$_3$) made of hydrogen bonded centrosymmetric dimers (HCO$_3^-)_2$ separated by K$^+$ entities \cite{TTO1,TTO2}. The coexistence of two configurations for protons was first regarded as statistical in nature \cite{BHT,EGS}. However, vibrational spectra (infrared, Raman, inelastic neutron scattering-INS) and neutron diffraction evidence the macroscopic quantum behaviour of protons and theoretical arguments suggest that this behaviour is intrinsic to the crystal state \cite{Fil3,IF,FCKeen,FCG2,FCG4}. 

In the present paper, we examine whether macroscopic states survive in a mixed crystal KH$_{p}$D$_{1-p}$CO$_3$, grown from a water solution composed of H$_2$O, HDO, D$_2$O molecules. Since there is no segregation of H and D atoms into separate domains or clusters, it might seem straightforward to suppose that the crystal is a mixture of HH, HD, DD dimers, with probabilities $p^2, 2p(1-p), (1-p)^2$, respectively \cite{TT,XHJT}. A random distribution of such dimers should destroy the translational invariance of the crystal, as well as---needless to say---the space group symmetry, and this should be evidenced by neutron diffraction. 

We report below neutron diffraction measurements showing that a mixed crystal ($p \approx 0.75$) is isomorphous to the pure analogues KHCO$_3$ and KDCO$_3$. This is at variance with any random distribution of H and D nuclei, so the concept of macroscopic states previously elaborated for pure crystals should hold for isotope mixtures, as well. 

The organization of this paper is as follows. The crystal structure of KHCO$_3$ at 300 K is presented in Sec. \ref{sec:1} and the theoretical framework for H or D macroscopic states is presented in Sec. \ref{sec:2}. In Sec. \ref{sec:3} we present evidences of macroscopic proton states obtained with neutron diffraction. Finally, some consequences of the $P2_1/a$ space group symmetry determined with neutron diffraction for the mixed crystal are discussed in Sec. \ref{sec:4}. 

\section{\label{sec:1}The crystal structure of KHCO$_3$ and KDCO$_3$}

\begin{figure}[!hbtp]
\begin{center}
\includegraphics[scale=.2]{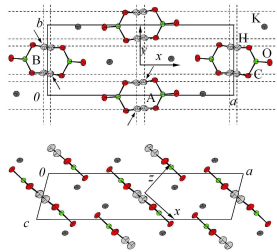}
\end{center}
\caption{\label{fig:1} Schematic view of the crystalline structure of KHCO$_3$ at 300 K. The arrows point to the sites occupied at low temperature. Dashed lines through protons are guides for the eyes. The ellipsoids represent 50\% of the probability density for nuclei.}
\end{figure}

The crystal of KHCO$_3$ is monoclinic, space group P$2_1/a$ (C$_{2h}^5$), with four equivalent entities per unit cell (Fig. \ref{fig:1}). Centrosymmetric dimers (HCO$_3^-)_2$ linked by moderately strong hydrogen bonds, with lengths $R_{\mathrm{OO}} \approx  2.58$ \AA, are well separated by K$^+$ ions. All dimers lie practically in (103) planes, hydrogen bonds are virtually parallel to each other, and all protons are crystallographically equivalent (indistinguishable). This crystal structure is unique to probing proton dynamics along directions $x,\ y,\ z,$ parallel to the stretching ($\nu$OH), the in-plane bending ($\delta$OH), and the out-of-plane bending ($\gamma$OH) coordinates, respectively. 

From 14 K to 300 K, there is no structural phase transition. The increase of the unit cell dimensions and of the hydrogen bond length are marginal. Only the population of proton sites changes significantly. Below $\approx 150$ K, all dimers are in a unique configuration, say $L$ (see arrows in Fig. \ref{fig:1}). At elevated temperatures, protons are progressively transferred along the hydrogen bonds to less favored sites (configuration $R$) at $\approx 0.6$ \AA\ from the main position. The center of symmetry is preserved and all proton sites remain indistinguishable. The population of the less favored site (or interconversion degree $\rho \approx 0.22$) is determined by an asymmetric double-well \cite{Fil2,FTP}, but there are controversies as to whether this leads to statistical disorder \cite{BHT,EGS} or quantum delocalization \cite{FCG2,FCG4,FCG3,Fil7}. 

The isomorphous $P2_1/a$ crystal of KDCO$_3$ is made of a very similar arrangement of centrosymmetric dimers. The hydrogen bond length is slightly longer ($R_{\mathrm{OO}} \approx  2.61$ \AA) and the population of the less favored site is smaller ($\rho \approx 0.12$) \cite{TTO2}. This suggests a slightly greater potential asymmetry for KDCO$_3$ than for KHCO$_3$ \cite{Fil2}. 

\section{\label{sec:2}Theory}

In this section we present the bases of the theoretical framework leading to macroscopic states for protons or deuterons \cite{FCG2,FCG4}.

\subsection{\label{sec:21}The adiabatic separation}

Within the framework of the Born-Oppenheimer approximation, the vibrational Hamiltonian can be partitioned as
\begin{equation}\label{eq:1}
\mathcal{H}_\mathrm{v} = \mathcal{H}_{\mathrm{H/D}} +\mathcal{H}_{\mathrm{at}}+ \mathcal{C}_{\mathrm{(H/D)at}},
\end{equation}
where $\mathcal{H}_{\mathrm{H/D}}$ and $\mathcal{H}_{\mathrm{at}}$ represent the sublattices of light nuclei (H$^+$ or D$^+$) and heavy atoms, respectively, while $\mathcal{C}_{\mathrm{(H/D)at}}$ couples the subsystems. For OHO or ODO bonds, coupling terms between OH/OD and O$\cdots$O degrees of freedom are rather large \cite{SZS,Novak}, hence beyond the framework of the perturbation theory. Two approaches, either semiclassical or quantum, are envisaged.

Semiclassical protons or deuterons are dimensionless particles moving across a potential hypersurface \cite{BVIT,BVT,GPNK,SFS2,TVL}. Complex trajectories involving heavy atom coordinates lead to mass renormalization, and to incoherent phonon-assisted tunnelling \cite{BHT,EGS,ST}. This approach is quite natural when the Born-Oppenheimer surface is known, but quantum effects can be severely underestimated. 

Alternatively, if the classical concept of ``trajectory'', totally alien to quantum mechanics, is abandoned, adiabatic separation of the two subsystems, namely $\mathcal{H}_{\mathrm{H/D}}$ and $\mathcal{H}_{\mathrm{at}}$, may lead to tractable models \cite{SZS,FCG2,GPNK,FRLL,W,MW}. Then, protons or deuterons in a definite eigen state should remain in the same state in the course of time, while heavy atoms oscillate slowly, in an adiabatic hyperpotential depending on the proton/deuteron state, through the coupling term. This separation is relevant for KHCO$_3$ and KDCO$_3$ because adiabatic potentials for different proton or deuteron states do not intersect each other. In fact, the separation is rigorously exact in the ground state, since protons or deuterons should remain in this state for ever, if there is no external perturbation. Then, protons are bare fermions, while deuterons are bosons, so quantum correlations should be different for the two crystals. 

\subsection{\label{sec:22}Macroscopic states}

Consider a crystal composed of very large numbers $N_\mathrm{a}$, $N_\mathrm{b}$, $N_\mathrm{c}$ ($\mathcal{N}=N_\mathrm{a}N_\mathrm{b}N_\mathrm{c}$) of unit cells labelled $j,k,l,$ along crystal axes $(a),$ $(b),$ $(c)$, respectively. The two dimers per unit cell are indexed as $j,k,l$ and $j',k,l$, respectively, with $j = j'$. For centrosymmetric dimers, there is no permanent dipolar interaction, so interdimer coupling terms and phonon dispersion are negligible \cite{FTP,IKSYBF,KIN}. The eigen states of the sublattice of protons can be therefore represented in a rather simple way with the basis sets of eigen states for isolated dimers 

A dimer (H1,H2) or (D1,D2) is modelled with coupled centrosymmetric collinear oscillators in three dimensions, along coordinates $\{\alpha_{1\mathrm{jkl}}\}$ and $\{\alpha_{2\mathrm{jkl}}\}$ ($\alpha =x,y,z$). The center of symmetry is at $\{\alpha_{0\mathrm{jkl}}\}$. The mass-conserving symmetry coordinates independent of $j, k,l,$ and their conjugated momenta, 
\begin{equation}\label{eq:2}
  \begin{array}{lc}
\alpha_{\mathrm{s}} = \displaystyle{\frac {1} {\sqrt{2}}\left(\alpha_{1} - \alpha_{2} + 2\alpha_0 \right)}, & P_{\mathrm{s}\alpha} = \displaystyle{\frac {1} {\sqrt{2}} \left( P_{1\alpha} - P_{2\alpha} \right)}, \\
    \alpha_{\mathrm{a}} = \displaystyle{\frac {1} {\sqrt{2}}\left(\alpha_{1} + \alpha_{2} \right)}, & P_{\mathrm{a}\alpha} = \displaystyle{\frac {1} {\sqrt{2}}\left( P_{1\alpha} + P_{2\alpha } \right)},
  \end{array}
\end{equation}
lead to uncoupled oscillators at frequencies $\hbar\omega_{\mathrm{s}\alpha}$ and $\hbar\omega_{\mathrm{a}\alpha}$, respectively, each with $m = 1$ amu. The difference $(\hbar\omega_{\mathrm{s}\alpha} - \hbar\omega_{\mathrm{a}\alpha})$ depends on the coupling term (say $\lambda_\alpha$). The wave functions, $\{\Psi_{\mathrm{njkl}}^\mathrm{a}( \alpha_{\mathrm{a}})\}$, $\{\Psi_{\mathrm{n'jkl}}^\mathrm{s}(\alpha_{\mathrm{s}} -\sqrt{2}\alpha_{0})\}$, cannot be factored into wave functions for individual particles, so there is no local information available for these entangled oscillators. Then, the wave functions of protons (fermions) or deuterons (bosons) are subject to the symmetrization postulate of quantum mechanics \cite{CTDL}.

\subsubsection{\label{sec:221}Protons}

The degenerate ground state of indistinguishable fermions must be antisymmetrized with respect to site permutation. For this purpose, the wave function is written as a linear combination 
\begin{equation}\label{eq:3}
\Theta_{0\mathrm{jkl}\pm } = \displaystyle{\frac{1} {\sqrt{2}}} \prod\limits_\alpha \Psi_{0\mathrm{jkl}}^\mathrm{a} (\alpha_{\mathrm{a}}) \left [ \Psi_{0\mathrm{jkl}}^\mathrm{s}(\alpha_{\mathrm{s}} -\sqrt{2}\alpha_{0}) \pm \Psi_{0\mathrm{jkl}}^\mathrm{s} (\alpha_{\mathrm{s}} +\sqrt{2}\alpha_{0}) \right ]
\end{equation}
and the antisymmetrized state vectors with singlet-like $|S\rangle$ or triplet-like $|T\rangle$ spin symmetry are: 
\begin{equation}\label{eq:4}
\begin{array}{rcl}
|0jkl+ \rangle \otimes |S\rangle & = & \big | \Theta_{0\mathrm{jkl}+} \rangle \otimes \displaystyle{\frac{1}{\sqrt{2}}} \left [ |\uparrow_1 \downarrow_2 \rangle - | \downarrow_1 \uparrow_2 \rangle \right ] ;\\
|0jkl- \rangle \otimes |T\rangle & = & | \Theta_{0\mathrm{jkl}-} \rangle \otimes \displaystyle{\frac{1}{\sqrt{3}}} \Big \{ | \uparrow_1 \uparrow_2 \rangle + | \downarrow_1 \downarrow_2 \rangle \\
& + & \displaystyle{\frac{1}{\sqrt{2}}} [ |\uparrow_1 \downarrow_2 \rangle + |\downarrow_1 \uparrow_2 \rangle ] \Big \} .
\end{array} 
\end{equation}
The oscillators are now entangled in position, momentum, and spin. In contrast to magnetic systems \cite{Cowley}, there is no level splitting, so the symmetry related entanglement is energy-free. It is also independent of $\lambda_\alpha$. Furthermore, as there is no significant exchange integral for protons separated by $\approx 2.2$ \AA\ \cite{FCou}, protons are not itinerant particles and there is no sizeable energy band width. 

Consider now the whole sublattice of protons. The spatial periodicity leads to collective dynamics and nonlocal observables in three dimensions. With the vibrational wave function for the unit cell $j,\ k,\ l,$ namely $\Xi_{0\mathrm{jkl}\tau } = \Theta_{0\mathrm{jkl}\tau } \pm \Theta_{0\mathrm{j'kl}\tau }$, where $\tau =$ ``$+$'' or ``$-$'' for singlet-like or triplet-like symmetry, respectively, trial phonon waves are written as 
\begin{equation}\label{eq:5}
\Xi_{0\tau} (\mathbf{k})= \displaystyle{\frac{1}{\sqrt {\mathcal{N}}}}  \sum\limits_{\mathrm{l} = 1}^{\mathrm{N_c}} \sum\limits_{\mathrm{k} = 1}^{\mathrm{N_b}} \sum\limits_{\mathrm{j} = 1}^{\mathrm{N_a}} \Xi_{0\mathrm{jkl}\tau} \exp(i\mathbf{k\cdot L}), 
\end{equation}
where $\mathbf{k}$ is the wave vector, $\mathbf{L}  = j \mathbf{a} + k \mathbf{b} + l \mathbf{c}$ is the lattice vector, $\mathbf{a}$, $\mathbf{b}$, $\mathbf{c}$, are the unit cell vectors. This equation could represent collective dynamics of (H,H) dimers if they were composed bosons in a crystal made of indistinguishable dimer entities (KHCO$_3)_2$. However, x-ray and neutron diffraction show that the structure is composed of monomer entities (KHCO$_3$) related to each other through symmetry operations. The probability density of each atom is equally distributed over all equivalent sites and, conversely, the probability density at each site includes contributions from all indistinguishable nuclei of the same kind. Consequently, antisymmetrization applied to the sublattice of indistinguishable fermions leads to 
\begin{equation}\label{eq:6}
\mathbf{k\cdot L} \equiv 0 \mathrm{\ modulo\ } 2\pi. 
\end{equation}
This means that there is no phonon (no elastic distortion) in the ground state. This symmetry-related ``super-rigidity'' \cite{FCG2} is totaly independent of any proton--proton interaction. The state vectors in three dimensions can be then written as: 
\begin{equation}\label{eq:7}
\begin{array}{c}
|\mathrm{H}_+\rangle = \left | \Xi_{0 +} (\mathbf{k = 0}) \right \rangle \otimes |S \rangle; \\
|\mathrm{H}_-\rangle = \left | \Xi_{0 -} (\mathbf{k = 0}) \right \rangle \otimes |T \rangle. 
\end{array}
\end{equation}
The wave functions $\Xi_{0 \tau} (\mathbf{k = 0})$ represent collective oscillations of the super-rigid lattice as a whole, with respect to the center of mass of the crystal. Finally, the ground state of the sublattice is a superposition state as:
\begin{equation}\label{eq:8}
\begin{array}{c}
\sqrt{\mathcal{N}} | \Xi_{0 +} (\mathbf{k = 0}) \rangle \otimes |S \rangle;\\ 
\sqrt{\mathcal{N}} | \Xi_{0 -} (\mathbf{k = 0}) \rangle \otimes |T \rangle. 
\end{array}
\end{equation}
This ground state is intrinsically steady against decoherence. Irradiation by plane waves (photons or neutrons) may single out some excited states. Entanglement in position and momentum is preserved, while the spin-symmetry and super-rigidity are destroyed. However, the spin-symmetry reappears automatically after decay to the ground state, presumably on the time-scale of proton dynamics. Consequently, disentanglement reaches a steady regime such that the amount of transitory disentangled states is determined by the ratio of density-of-states for the surroundings (atmosphere, external radiations...) and for the crystal, respectively. This ratio is so small that disentangled states are too few to be observed. Nevertheless, they allow the super-rigid sublattice to be at thermal equilibrium with the surroundings, despite the lack of internal dynamics. 

The main source of disentanglement is the thermal population of excited proton states. However, even at room temperature, the thermal population of the first excited state ($< 1\%$ for $\gamma$OH $\approx 1000$ \cm) is of little impact to measurements. 

\subsubsection{\label{sec:222}Deuterons}

For the sublattice of bosons in the isomorphic crystal of KDCO$_3$, Eqs (\ref{eq:3}), (\ref{eq:4}) and (\ref{eq:6}) are not relevant. There is neither spin-symmetry nor super-rigidity. Dynamics are represented with symmetry coordinates (\ref{eq:2}) leading to wave functions 
\begin{equation}\label{eq:9}
\Theta_{0\mathrm{jkl}} = \prod\limits_\alpha \Psi_{0\mathrm{jkl}}^\mathrm{a} (\alpha_{\mathrm{a}}) \Psi_{0\mathrm{jkl}}^\mathrm{s}(\alpha_{\mathrm{s}} -\sqrt{2}\alpha_{0}) ,
\end{equation}
and phonon states 
\begin{equation}\label{eq:10}
|\mathrm{D}_{\mathbf{k}\pm}\rangle= \displaystyle{\frac{1}{\sqrt {\mathcal{N}}}}  \sum\limits_{\mathrm{l} = 1}^{\mathrm{N_c}} \sum\limits_{\mathrm{k} = 1}^{\mathrm{N_b}} \sum\limits_{\mathrm{j} = 1}^{\mathrm{N_a}} |\Xi_{0\mathrm{jkl}\pm}\rangle \exp(i\mathbf{k\cdot L}), 
\end{equation}
where $\Xi_{0\mathrm{jkl}\pm} = \Theta_{0\mathrm{jkl}} \pm \Theta_{0\mathrm{j'kl}}$. Needless to say, the H and D nuclei have the same number of degrees of freedom (12 per unit cell), but the symmetrization postulate shrinks the size of the allowed Hilbert space from $\sim 12^\mathcal{N}$ for bosons to $\sim 12\mathcal{N}$ for fermions. 

\section{\label{sec:3}Probing quantum entanglement with neutrons}

Neutrons (spin $1/2$) are unique to probing the spin-symmetry of macroscopic proton states (\ref{eq:7}). However, quantum entanglement is extremely fragile, as it is not stabilized by any energy. Only ``noninvasive'' experiments, free of measurement-induced decoherence, are appropriate \cite{LG}. For neutron scattering, this means (i) no energy transfer (ii) no spin-flip and (iii) particular values of the momentum transfer vector $\Q$. ($\Q = \mathbf{k}_\mathrm{i} - \mathbf{k}_\mathrm{f}$, where  $\mathbf{k}_\mathrm{i}$ and $\mathbf{k}_\mathrm{f}$ are the initial and final wave vectors, respectively.) 

\begin{figure}[!htbp]
\includegraphics[angle=0.,scale=0.35]{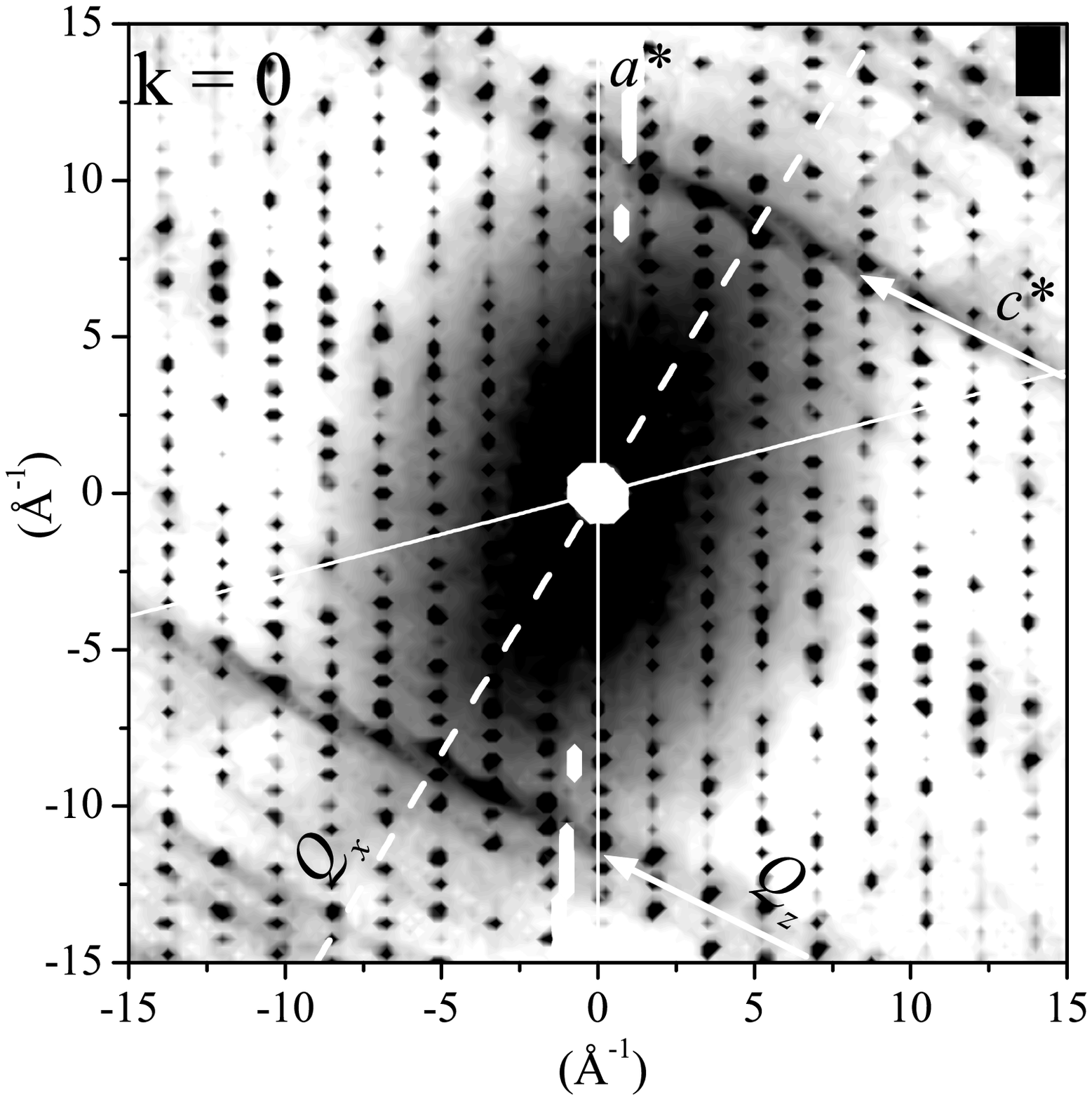}
\includegraphics[angle=0.,scale=0.35]{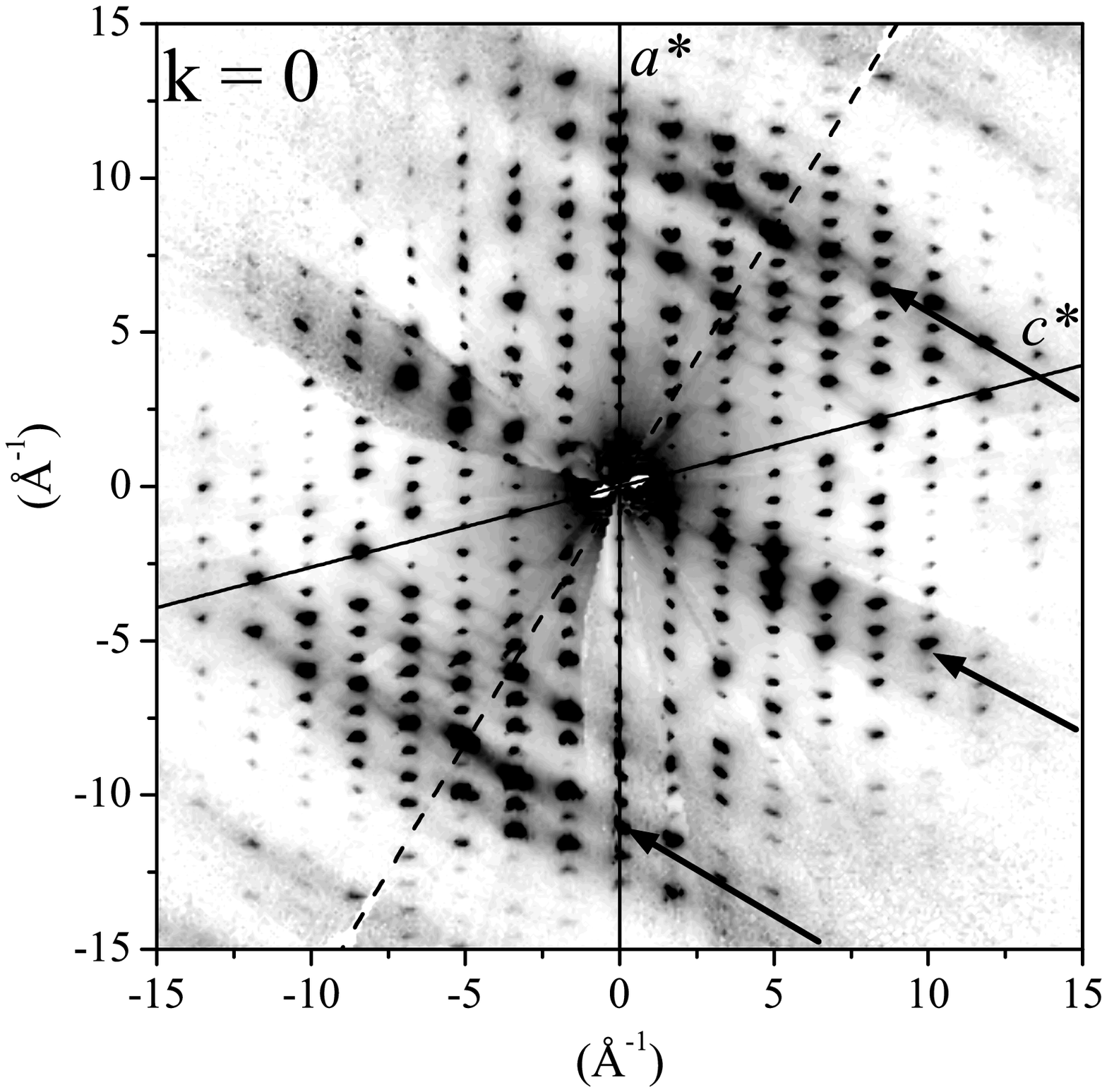}
\caption{\label{fig:2} Diffraction patterns of KHCO$_3$ at 30 K (left) and 300 K (right) in the ($a^*, c^*$) reciprocal plane at k = 0, after Ref. \cite{FCG2}. The arrows emphasize ridges of intensity parallel to $Q_\mathrm{z}$ and perpendicular to the dimer plane (dash lines along $Q_\mathrm{x}$).}
\end{figure}

Noninvasive diffraction events preserving the super-rigidity occur when components of the momentum transfer along proton coordinates ($Q_\mathrm{x}$, $\ Q_\mathrm{y}$, $\ Q_\mathrm{z}$) match a node of the reciprocal sublattice of protons, so the momentum transfer is the same at all sites. Then, neutrons probe super-rigid states without any induced distortion. The only information conveyed by such events is the perfect periodicity of the sublattice, so the Debye-Waller factor is equal to unity at any temperature. In addition, thanks to the spin-symmetry, the scattered intensity is proportional to the total cross-section $\sigma_\mathrm{H} \approx 82.0$ b \cite{FCG2,SWL}. Otherwise, if the matching condition is not realized, (i) the spin-symmetry is destroyed, (ii) the intensity scattered by protons is proportional to the coherent cross-section $\sigma_{\mathrm{Hc}} \approx 1.76$ b, (iii) Bragg-peaks are depressed by the usual Debye-Waller factor for non-rigid lattices. Quantum correlations can be therefore clearly observed, thanks to the dramatic enhancement factor $\sigma_{\mathrm{H}} / \sigma_{\mathrm{Hc}} \approx 45$. Furthermore, the intensity scattered by heavy atoms, proportional to $\sigma_{\mathrm{cKCO_3}} \approx 27.7$ b, is depressed by the Debye-Waller factor: $\exp-2W_{\mathrm{KCO_3}}(\Q)$. Therefore, the contribution at large $\Q$-values is much weaker than that of the entangled sublattice and the contrast of intensity is enhanced at elevated temperatures. 

The dashed lines in Fig. \ref{fig:1} show that proton sites are aligned along $x$ and $y$, but not along $z$. Consequently, the noninvasive condition can be realized for $Q_\mathrm{x}$ and $\ Q_\mathrm{y}$, whereas $Q_\mathrm{z}$ does not coincide, strictly speaking, with a nod of the reciprocal lattice of protons. The diffraction pattern is composed of rods of diffuse scattering parallel to $Q_\mathrm{z}$, cigar-like shaped by the Debye-Waller factor. A detailed analysis of the scattering function reveals a complex pattern of rods of intensity, in full agreement with observations \cite{FCG2,FCG4}. For example, Fig. \ref{fig:2} shows slices of the diffraction patterns of KHCO$_3$ at 30 and 300 K, in the $(a^*,c^*)$ reciprocal plane, at $Q_\mathrm{y} = 0$ (k $= Q_y/b^* = 0$). The cigar-like shaped rods are observed at $Q_\mathrm{x} = 0$ and $\pm (10.00 \pm 0.25)$ \A, in accordance with the distance of $\approx 0.6$ \AA\ between double lines of proton. These rods are evidences of macroscopic quantum correlations in the sublattice of protons. As anticipated from (\ref{eq:10}), these rods are not observed for KDCO$_3$ \cite{FCKeen}. 

\begin{figure}[!hbtp]
\begin{center}
\includegraphics[scale=0.2, angle=0]{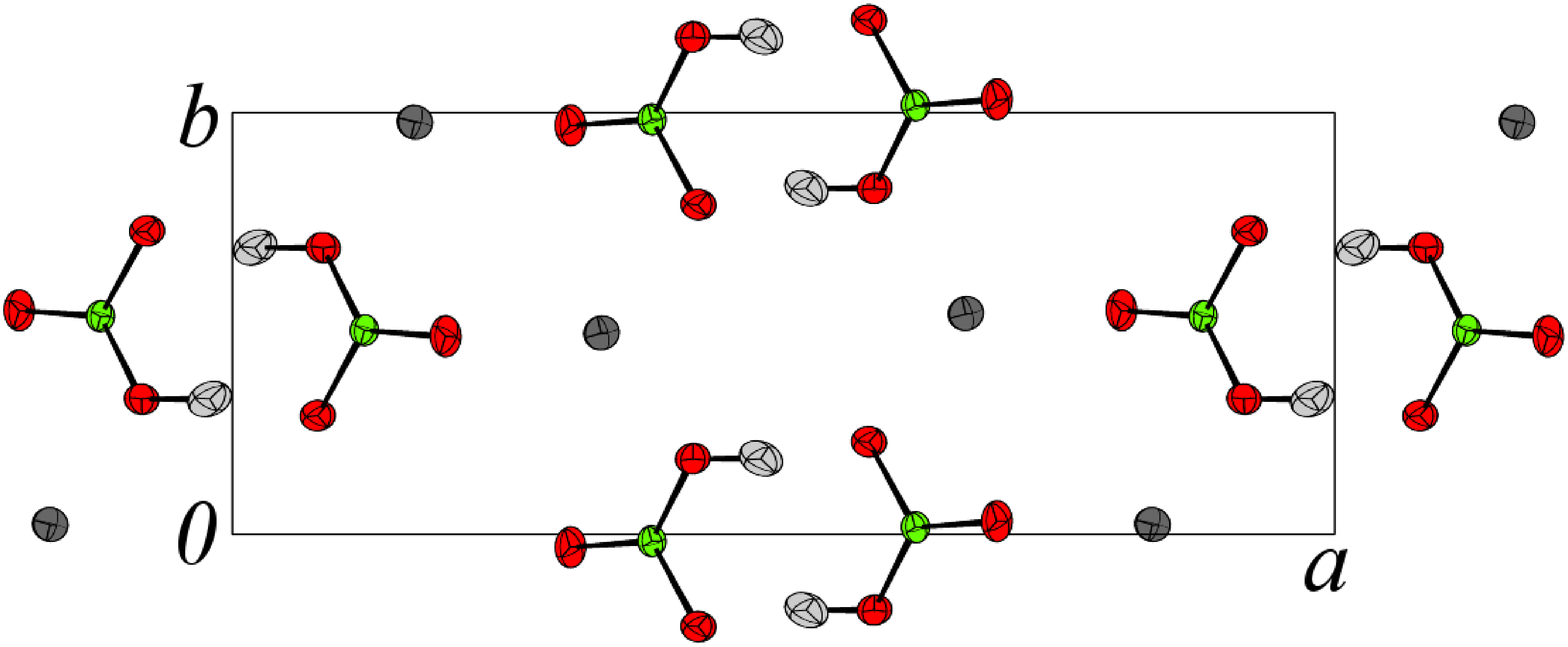}
\makebox[\textwidth]{\ }\\
\includegraphics[scale=0.2, angle=0]{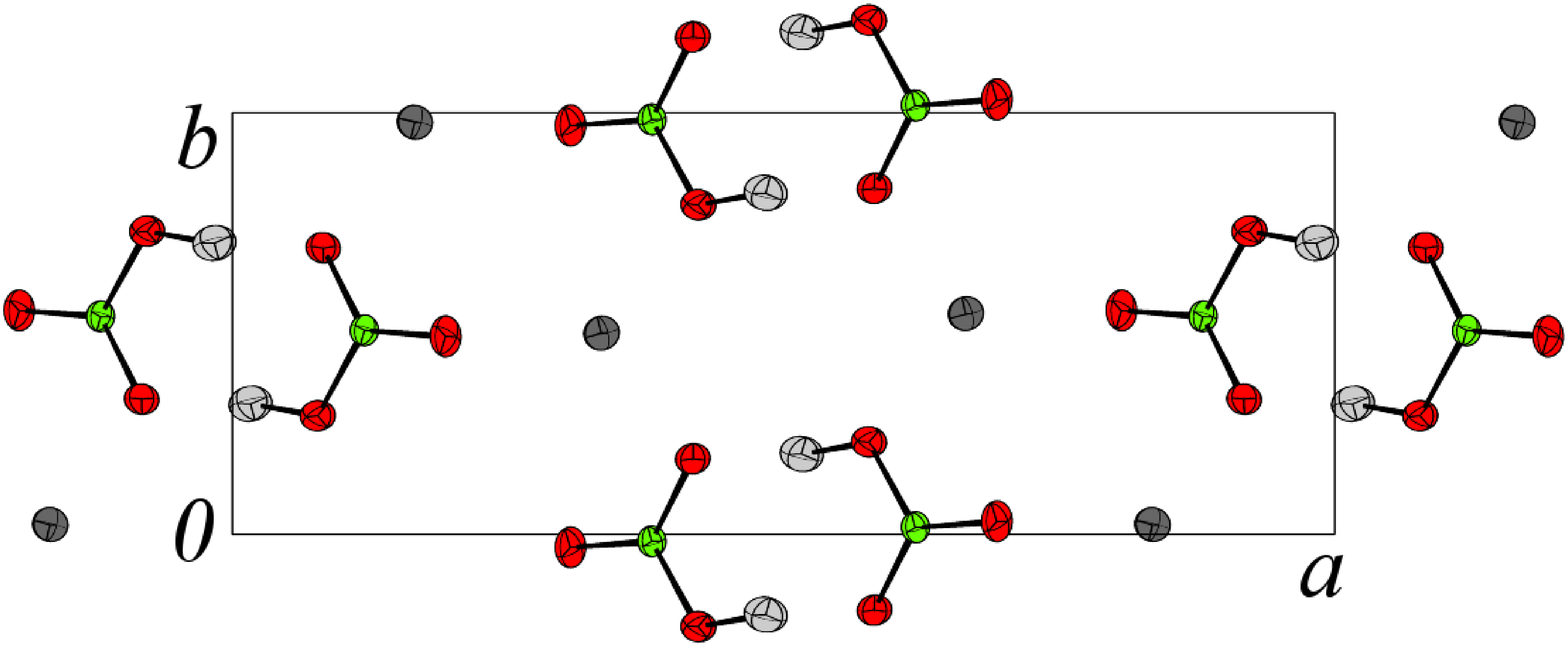}
\end{center}
\caption{\label{fig:3} Projections on the ($a,b$) plane of the $L$ (bottom) and $R$ (top) configurations. }
\end{figure}

At low temperatures, the rods are partially hidden by the anisotropic continuum, centered at $\Q =0$, due to incoherent scattering. At 300 K, this continuum is dramatically depressed by the Debye-Waller factor, while diffraction by the super-rigid sublattice is largely unaffected. Quite remarkably, quantum correlations are not destroyed by interconversion. The sublattice is therefore a superposition of macroscopic states, $|L\rangle$ and $|R\rangle$, corresponding to the structures sketched in Fig. \ref{fig:3}. 

Theory and experiments show that protons (deuterons) in the crystal field are not individual particles possessing properties in their own rights. Instead of that, they merge into macroscopic states. The theory suggests that such matter fields, intrinsic to translational invariance, should occur in many crystals. However, the diffraction patterns in Fig. \ref{fig:2} are easy to interpret because protons are indistinguishable and hydrogen bonds are aligned (Fig. \ref{fig:1}). For more complex structures, seeking rods of diffuse scattering in diffraction patterns could be problematic. In the next section, we propose a complementary approach to evidence the nonlocal nature of protons and deuterons. 

\section{\label{sec:4}Isotope mixtures}

In the present section we examine whether macroscopic states exist in isotope mixtures KH$_p$D$_{1-p}$CO$_3$. We anticipate that any superposition of (\ref{eq:7}) and (\ref{eq:10}) should preserve the $P2_1/a$ space group symmetry of the crystal, whereas any statistical distribution of isotopes, here or there, should destroy the translational invariance. Because neutrons are scattered by nuclei, the diffraction pattern should discriminate between these two possibilities. By contrast, the $P2_1/a$ symmetry evidenced by x-ray would not exclude, per se, a random distribution of isotopes, for they have identical electronic structures. 

A nearly cubic specimen ($3\times 3\times 3$ mm$^3$) was cut from a large crystal grown by cooling slowly a saturated solution in a mixture of H$_2$O ($\approx 75\%$) and D$_2$O ($\approx 25\%$). Measurements were conducted with the four-circle diffractometer 5C2 at the Orph\'{e}e reactor (Laboratoire L\'{e}on-Brillouin) \cite{LLB}. The crystal wrapped in aluminum was maintained at $(300 \pm 1)$ K. Data reduction was carried out with CRYSTALS \cite{WPCBC,CRYSTALS}. 

\begin{table}[!hbtp]
\caption{\label{tab:1} Unit cell parameters in \AA\ and $^\circ$ units determined with single-crystal neutron diffraction at 300 K. $\lambda$ = 0.8305 \AA, space group $P 2_1/a$. The variance for the last digit is given in parentheses.}
\begin{center}
\begin{tabular}{lll}
\hline
 & KH$_{0.75}$D$_{0.25}$CO$_3$ & KHCO$_3$ \cite{FCG2} \\
\hline
$a$(\AA) & 15.180(1) & 15.180(1) \\
$b$(\AA) & 5.620(1)  & 5.620(4) \\
$c$(\AA) & 3.710(1)  & 3.710(4) \\
$\beta$  & 104.67(1)$^\circ$ & 104.67(5)$^\circ$ \\
\hline
\end{tabular}
\end{center}
\end{table}

A preliminary inspection of intensities at the absent reflections ($0 k 0$ for $k \neq 2n$ and $h 0 l$ for $h \neq 2n$) confirms the $P2_1/a$ space group assignment. The unit cell parameters in Table \ref{tab:1} are identical to those of KHCO$_3$. A complete determination of the positional and thermal parameters is currently in progress. 

The $P2_1/a$ symmetry is at variance with a distribution of H and D nuclei at random. It is consistent with the existence of four sublattices, H$_L$, H$_R$, D$_L$, D$_R$, analogous to those sketched in Fig. \ref{fig:3}, such as: 
\begin{equation}\label{eq:15}
\begin{array}{l}
\mathrm{HD}(p, \rho_{H}, \rho_{D}, ) = \\
p [(1-\rho_{H})\mathrm{H}_L + \rho_{H} \mathrm{H}_R] + (1-p) [(1-\rho_{D})\mathrm{D}_L + \rho_{D} \mathrm{D}_R]. 
\end{array}\end{equation}

With a diffractometer based at a reactor source one measures a limited range of the reciprocal space, typically $|\Q| \lessapprox 5$ \A. The matching condition for noninvasive measurements is realized only at $Q_\mathrm{x} = Q_\mathrm{y} =0$, so for almost all measured Bragg-peaks, the spin symmetry is destroyed, the cross-section is $\sigma_\mathrm{Hc}$, and the sublattice of protons behaves exactly the same as a non-rigid lattice of bosons (\ref{eq:10}). The four sublattices are therefore consistent with a superposition of separable macroscopic states such as: $|\mathrm{H}_{\mathbf{k}\pm}\rangle_L$, $|\mathrm{H}_{\mathbf{k}\pm}\rangle_R$, $|\mathrm{D}_{\mathbf{k}\pm} \rangle_L$, $|\mathrm{D}_{\mathbf{k}\pm} \rangle_R$. There is no direct information as to whether quantum correlations for proton states hold in the mixed sublattice. This point deserves further investigations. 

\section{Conclusion}

The $P2_1/a$ space group symmetry of the KH$_{p}$D$_{1-p}$CO$_3$ ($p \approx 0.75$) crystal determined with neutron diffraction shows that H and D nuclei are not individual particles located at definite sites and mutually exclusive. Wave-like protons and deuterons are equally distributed over all indistinguishable sites and, conversely, each site is occupied by all wave-like H and D nuclei with total occupancies $p$ and $(1-p)$, respectively. This nonlocal behaviour allows the sublattice H$_p$D$_{1-p}$ to match the symmetry of the electronic structure that is unaffected by isotope substitution. The superposition of macroscopic states sounds as weird as a Cat in a superposition of $|\mathrm{alive}\rangle$ and $|\mathrm{dead}\rangle$ states. 

To the question ``Where are protons and deuterons?'', our answer is that the particle-like representation, implicitly meant by ``Where'', is irrelevant. Only the wave-like, or matter-field, representation is permitted by the crystal structure. The theory is based on fundamental principles of quantum mechanics and it is free of any ad hoc hypothesis or parameter. The nonlocal nature of protons and deuterons could be therefore of general significance. Neutron diffraction by hydrogen bonded crystals with partial isotope substitution could be a well suited and powerful approach to study nonlocal behaviors.

\end{document}